\documentclass[aps,prb,twocolumn,superscriptaddress]{revtex4-1}
\usepackage{amsmath}
\usepackage{graphicx}
\usepackage{hyperref}
\begin{document}

%\preprint{}

\title{Nonlocal Exchange Interactions in Strongly Correlated Electron Systems}


\author{Edin Kapetanovi\'{c}}
\affiliation{Institut f\"{u}r Theoretische Physik, Universit\"{a}t Bremen, Otto-Hahn-Allee 1, D-28359 Bremen}
\affiliation{Bremen Center for Computational Materials Science, Universit\"{a}t Bremen, Am Fallturm 1a, D-28359 Bremen}
%\thanks{}
\author{Malte Sch\"{u}ler}
\affiliation{Institut f\"{u}r Theoretische Physik, Universit\"{a}t Bremen, Otto-Hahn-Allee 1, D-28359 Bremen}
\affiliation{Bremen Center for Computational Materials Science, Universit\"{a}t Bremen, Am Fallturm 1a, D-28359 Bremen}
\author{Gerd Czycholl}
\affiliation{Institut f\"{u}r Theoretische Physik, Universit\"{a}t Bremen, Otto-Hahn-Allee 1, D-28359 Bremen}
\author{Tim O. Wehling}
\affiliation{Institut f\"{u}r Theoretische Physik, Universit\"{a}t Bremen, Otto-Hahn-Allee 1, D-28359 Bremen}
\affiliation{Bremen Center for Computational Materials Science, Universit\"{a}t Bremen, Am Fallturm 1a, D-28359 Bremen}

\date{\today}

\begin{abstract}
%Contrary to the nonlocal Coulomb repulsion which is included in many studies of extended Hubbard models, very little is known about the influence of the nonlocal exchange interaction. In order to include the exchange in the framework of a Hubbard model, we introduce a novel, variational approach which is capable of capturing correlations, and apply it to a half-filled square lattice which we simulate with Determinantal Quantum Monte Carlo. In contrast to a simple mean-field approach which predicts a first order transition between a Néel-state and ferromagnetic order, we find smooth transitions between areas in the phase diagram where antiferromagnetic and ferromagnetic correlations dominate.
We study the influence of ferromagnetic nonlocal exchange on correlated electrons in terms of a $SU(2)$-Hubbard-Heisenberg model and address the interplay of on-site interaction induced local moment formation and the competition of ferromagnetic direct and antiferromagnetic kinetic exchange interactions. In order to simulate thermodynamic properties of the system in a way that largely accounts for the on-site interaction driven correlations in the system, we advance the correlated variational scheme introduced in [M. Sch\"{u}ler et al., Phys. Rev. Lett. 111, 036601 (2013)] to account for explicitily symmetry broken electronic phases by introducing an auxiliary magnetic field. After benchmarking the method against exact solutions of a finite system, we study the $SU(2)$ Hubbard-Heisenberg model on a square lattice. We obtain the $U$-$J$ finite temperature phase diagram of a $SU(2)$-Hubbard-Heisenberg model within the correlated variational approach and compare to static mean field theory. While the generalized variational principle and static mean field theory yield transitions from dominant ferromagnetic to antiferromagnetic correlations in similar regions of the phase diagram, we find that the nature of the associated phase tranistions differs between the two approaches. The fluctuations accounted for in the generalized variational approach render the transitions continuous, while static mean field theory predicts discontinuous transitions between ferro- and antiferromagnetically ordered states.
\end{abstract}

%\keywords{}

\maketitle

\section{Introduction}
One of the most intensively studied and most fundamental models for the description of correlated electrons on a crystal lattice is the Hubbard model \cite{Hubbard1,Hubbard2,Hubbard3,GutzwillerHubbardModel1963,KanamoriHubbardModel1963}. The central point of this model is to neglect all interactions aside from the local Coulomb repulsion between two electrons occupying the same lattice site. While the approximation of purely local interaction can pose a drastical simplification, the model is still capable of describing a wide range of phenomena from Mott-Hubbard metal-insulator transitions to unconventional superconductivity. This is why the model became a key for understanding the competition between itinerancy and localization due to interactions.%, which makes it subject to ongoing studies, with modern methods such as Dynamical Mean Field Theory (DMFT) \cite{DMFTVollhardt,DMFTGeorges}, Determinantal Quantum Monte Carlo (DQMC) \cite{DQMCIntroduction} and the Dynamical Cluster Approximation (DCA) \cite{DCAHettler1998}.
Several recently emerging quantum materials challenged the Hubbard model paradigm and realize correlated electron physics likely governed by more complex interaction patterns.

First, in low-dimensional and insulating systems, the neglected nonlocal Coulomb interactions play a significant role due to reduced screening, and the Hubbard model can fail to provide an adequate description. It is well known by now that nonlocal Coulomb repulsion in the form of so-called density-density terms can drive the system towards a charge density wave (CDW) \cite{GullCDW2DHubbard,BariCDW,Zhang2DExtendedHubbard},  effectively screen the local interaction \cite{OptHubbardModelsMalte}, influence possibly existing tendencies towards superconductivity \cite{ExtendedHubbardSuperconductivity1,ExtendedHubbardSuperconductivity2,ExtendedHubbardSuperconductivity3} and lead to a renormalization of Fermi velocities \cite{FermiVelocity1,FermiVelocity2}.

In systems like twisted bilayer graphene \cite{Cao2018_1,Cao2018_2} or other twisted 2D materials \cite{Wu_TopologicalInsulators_TwistedTMDCs}, general four fermion interactions are likely steering the low energy electrons \cite{Koshino_EHM_TBG} due to the intricate real space patterns of the low energy electronic Wannier functions. Currently, very little is known about the effects of general non-local four fermion interaction terms on electronic correlation phenomena. Among these are effects of e.g. correlated hopping terms and importantly non-local exchange interactions J.

Traditionally, $J$ has been neglected based on the smallness of the exchange integral ($J \sim 1/40 \ \mathrm{eV}$ for 3$d$-electrons) in comparison to the on-site repulsion ($U \sim 10 \ \mathrm{eV}$ \cite{Hubbard1}). However, this argumentation can be misleading. Generally, in the strong coupling case with $U \gg t$ exceeding the hopping $t$ by far, $J$ competes against the kinetic exchange given by $-4t^2/U$, which can be orders of magnitude smaller than the original $U$. For synthetic quantum materials like twisted bilayer graphene, where the electron correlations emerge beyond the atomic scale, it is very unclear why any estimates made for 3$d$-electron materials should transfer to this case.

In this paper, we advance a theoretical approach to account for interaction terms beyond the on-site Hubbard paradigm. We consider the explicit example of the Hubbard-Heisenberg model, which supplements the Hubbard model with non-local exchange terms, and analyze the interplay of on-site repulsion and non-local exchange effects. This problem has so-far been addressed in two different limits: First, $SU(N)$ generalizations of the Hubbard-Heisenberg model have been studied in the large-$N$ limit \cite{AffleckHubbardHeisenberg1,AffleckHubbardHeisenberg2,ChungHubbardHeisenbergTriangularLattice}.  Secondly, the important $SU(2)$-case has been studied in terms of Hartree-Fock mean-field theory (MFT) \cite{HubbardHeisenbergMeanFieldZeroT,CzartExtendedHubbardMeanField}. Here, we develop a variational approach to study the impact of quantum and thermal fluctuations on the interplay of local moment formation with ferromagnetic and antiferromagnetic spin correlations in the $SU(2)$ Hubbard-Heisenberg model.

The paper is structured as follows: In Sec.(\ref{ssec:Model}) we introduce the Hubbard-Heisenberg model. Sec. (\ref{ssec:VariationalPrinciple}) explains the generalization of the variational approach from \cite{OptHubbardModelsMalte} to account for explicitly symmetry broken phases and non-local exchange interactions: We map the Hubbard-Heisenberg Hamiltonian to an auxiliary Hamiltonian, which includes a renormalized Hubbard interaction and allows for breaking of the $SU(2)$ spin symmetry by including an effective, external magnetic field in $z$-direction. In Sec.(\ref{ssec:MonteCarlo}) we then give the computational details of the simulations of this auxiliary system performed with Determinantal Quantum Monte Carlo (DQMC).

In Sec.(\ref{ssec:Foursite}) we compare the exact solution of a 4-site Hubbard-Heisenberg cluster to approximate solutions from the variational approach developed here and to MFT for benchmarking purposes. Sec.(\ref{ssec:SquareLattice}) discusses the phase diagram and thermodynamic properties of the $SU(2)$ Hubbard-Heisenberg model on a square lattice. The phase diagrams obtained with the generalized variational principle and MFT are qualitatively similar. We find, however, that the fluctuations accounted for in the generalized variational approach render the transitions between ferro- and antiferromagnetically correlated states continuous, while MFT predicts discontinuous transitions. Furthermore, we illustrate and discuss the non-monotonous influence of a small, direct exchange, on correlation functions such as the double occupancy.
\section{Methods}\label{sec:Methods}

\subsection{The Model Hamiltonian}\label{ssec:Model}
Consider an extended Hubbard Model for electrons on a lattice which includes nonlocal interactions:
\begin{equation}
\begin{split}
	H = \sum_{i,j,\sigma} t_{ij} c_{i\sigma}^{\dagger} c_{j\sigma}^{\vphantom{\dagger}} + U \sum_i n_{i\uparrow} n_{i\downarrow}\\ \quad +\frac{1}{2} \sum_{i\neq j,\sigma,\sigma '} V_{ij} n_{i\sigma} n_{j\sigma '} - \frac{1}{2} \sum_{i\neq j}J_{ij} \vec{S}_i \cdot \vec{S}_j
\end{split}
\end{equation}
Here, $c_{i\sigma}^{\dagger}$ and $c_{i\sigma}^{\vphantom{\dagger}}$ denote the creation and annihilation operators for an electron in a Wannier state on site $i$ with the spin $\sigma$. The $t_{ij}$ contain the hopping matrix elements and the on-site energies. $U$ is the on-site interaction strength, while $V_{ij}$ and $J_{ij}$ are the nonlocal Coulomb repulsion and the exchange interaction, respectively. $n_{i\sigma} = c_{i\sigma}^{\dagger} c_{i\sigma}^{\vphantom{\dagger}}$ denotes the occupation number operator, while $\vec{S}_i$ represents the spin operator.

In previous papers \cite{OptHubbardModelsMalte,MalteFirstOrderMITScreening}, it was shown that the nonlocal Coulomb repulsion can be included in an effective Hubbard model with local interactions only, by renormalizing the on-site repulsion. While this mapping is an approximation, especially regarding the charge correlations, it works well for describing the spin dynamics since the $V$-term couples to the charge degrees of freedom. The nonlocal exchange, which we focus on, however, couples directly to the spin degrees of freedom. Thus, in this work, we assume that the nonlocal repulsion has already been absorbed into an effective Hubbard-$U$, and neglect the $V$-terms. We focus on a one-band model with next-neighbor hopping and interactions on a square lattice, which corresponds to the following Hubbard-Heisenberg-Hamiltonian:
\begin{equation}
\begin{split}
H = - t \sum_{\langle i,j \rangle , \sigma } \left(c_{i\sigma}^{\dagger} c_{j\sigma}^{\vphantom{\dagger}} + h.c. \right) + U \sum_i n_{i\uparrow} n_{i\downarrow} - J \sum_{\langle i,j \rangle} \vec{S}_i \cdot \vec{S}_j \label{eqn:OriginalHamiltonian}
\end{split}
\end{equation}
Here, the $\langle i,j \rangle$ denote pairs of nearest-neighbor sites.
\subsection{The Variational Principle}\label{ssec:VariationalPrinciple}
Our goal is to obtain an approximation to thermodynamic properties of the system defined in Eq. (2) by mapping it onto a simpler auxiliary system which is easier to handle. More precisely, we want to describe the properties of $H$ by mapping it onto an effective Hamiltonian $\tilde{H}$:
\begin{equation}
	\tilde{H} = - t \sum_{\langle i,j \rangle , \sigma } \left(c_{i\sigma}^{\dagger} c_{j\sigma}^{\vphantom{\dagger}} + h.c. \right) + \tilde{U} \sum_i n_{i\uparrow} n_{i\downarrow} - \tilde{B} \sum_i S_i^z \label{eqn:EffectiveHamiltonian}
\end{equation}
The effective, magnetic field $\tilde{B}$ has been introduced in order to implement ferromagnetic correlations, which stem from the Heisenberg-term in $H$ and cannot be captured in the framework of a simple half-filled Hubbard model. Keeping a renormalized on-site interaction $\tilde{U}$ allows this auxiliary system to capture correlations which go beyond Hartree-Fock theory. As shown in Appendix (\ref{sec:AppdxMeanField}), setting $\tilde{U} = 0$ is indeed equivalent to a mean-field description. A rationale behind introducing the auxiliary magnetic field $\tilde{B}$ is the following: the Hubbard model with on-site interactions has different low energy states \textit{close} to the ground state. The auxiliary field $\tilde B$ lowers those with desirable spin polarization in energy to achieve an optimized description of the full system.

For the mapping of the Hubbard Heisenberg model, Eq.(\ref{eqn:OriginalHamiltonian}), to the auxiliary Hubbard model in an external magnetic field, Eq.(\ref{eqn:EffectiveHamiltonian}), we make use of the Peierls-Feynman-Bogoliubov Variational principle \cite{PeierlsVariationalPrinciple,FeynmanVariationalPrinciple,BogoliubovVariationalPrinciple}.  $( \tilde{U}, \tilde{B})$ are variational parameters which are chosen so that the density operator $\rho_{\tilde{H}}$ of the auxiliary system $\tilde{H}$ approximates the real density operator $\rho_H$ as good as possible. In order to do so, we minimize the following expression with respect to the parameters $(\tilde{U},\tilde{B})$:

\begin{equation}
	\Phi_H \leq \tilde{\Phi} =  \langle H - \tilde{H} \rangle_{\tilde{H}} + \Phi_{\tilde{H}} \label{eqn:VariationGeneral}
\end{equation}

$\Phi_H = - \frac{1}{\beta} \ln Z_H$ with $Z_H = \textrm{Tr} (e^{-\beta H})$ being the partition function denotes the grand canonical potential\footnote{free energy in the canonical case} of the original Hamiltonian $H$ while $\Phi_{\tilde{H}}$ is the grand canonical potential of the auxiliary system. $\langle A \rangle_{\tilde{H}} = \frac{1}{Z_{\tilde{H}}} \textrm{Tr} (A e^{-\beta \tilde{H}})$ expresses the expectation value of an operator $A$ evaluated with the thermodynamic density operator of the effective Hamiltonian $\tilde{H}$. Evaluating the expression for $\tilde{\Phi}$ in Eq.(\ref{eqn:VariationGeneral}) leads to:

\begin{equation}
\begin{split}
	\tilde{\Phi} = \left(U -  \tilde{U}\right) \sum_i \langle n_{i\uparrow} n_{i\downarrow} \rangle_{\tilde{H}} - J \sum_{\langle i,j \rangle} \langle \vec{S}_i \cdot \vec{S}_j \rangle_{\tilde{H}} \\  + \tilde{B} \sum_i \langle S_i^z \rangle_{\tilde{H}} + \Phi_{\tilde{H}}
\end{split} \label{eq:FinalVariationalEquation}
\end{equation}

\subsection{Quantum Monte Carlo: Computational Details}\label{ssec:MonteCarlo}
We solve the effective Hamiltonian $\tilde{H}$ for different $(\tilde{U},\tilde{B})$ on a square lattice by performing Determinantal Quantum Monte Carlo (DQMC \cite{DQMCIntroduction}, QUEST code \cite{QUESTLink}) simulations. The raw data obtained from the simulation is available on \cite{DQMCRawData}. We restrict our calculations to the half-filled case for which no sign problem exists even for $\tilde{B} \neq 0$. The temperature for all simulations presented in this work is set to $\beta t = 10$, which means that for a square lattice, the thermal energy is on the order of $1/40$ of the free system's bandwidth. This temperature is cold enough to observe correlations and to capture the interesting phase transitions, particularly the metal-insulator transition \cite{SchaeferFateOfMottHubbardTransition}. In order to deal with finite size effects, we performed calculations for different system sizes (i.e. 4x4, 6x6, 8x8, 10x10, 12x12) and extrapolate to the $N \rightarrow \infty$ limit as described in \cite{MalteFirstOrderMITScreening}. 

In order to achieve good qualitative results, a rough estimate is to choose a discretization $\Delta \tau \sim \sqrt{0.125/U}$ for the Trotter-Suzuki decomposition \cite{DQMCIntroDosSantos}. For $U = 6$, which is the highest value that we use for the mapping, this leads to $\Delta \tau \approx 0.144$. In order to minimize the remaining Trotter-error, instead of simulating the system only at $\Delta \tau = 0.1$, we also simulate $\Delta \tau = 0.2$. This allows for an extrapolation $\Delta \tau \rightarrow 0$, since the Trotter error is known to scale with $\mathcal{O}(\Delta \tau ^2)$. In Appendix (\ref{sec:AppdxTrotter}), we provide an example which justifies the extrapolation with only two data points.

For each data point for $(\tilde{U},\tilde{B})$ with 28 data points $\tilde{U}/t = 0-6$ and 48 data points for $\tilde{B}/t = 0-4$, the simulation is run with 10000 warmup sweeps and 30000 measurement sweeps. In order to significantly reduce the Monte Carlo noise, we smooth the data with a two-dimensional Savitzky-Golay filter, as further explained in Appendix (\ref{ssec:AppdxErrorsDQMC}).

While the double occupancies and the spin-related expectation values appearing in Eq.(\ref{eq:FinalVariationalEquation}) can be directly measured within DQMC, determining the grand canonical potential $\Phi_{\tilde{H}}$ of the effective system requires a coupling constant integration. Since 
\begin{equation*}
\begin{split}
	- \frac{\partial \Phi_{\tilde{H}}}{\partial \tilde{B}} = \sum_i \langle S_i^z \rangle_{\tilde{H}} \\
	\frac{\partial \Phi_{\tilde{H}}}{\partial \tilde{U}} = \sum_i \langle n_{i\uparrow} n_{i\downarrow} \rangle_{\tilde{H}}
\end{split}
\end{equation*} 
$\Phi_{\tilde{H}}$ can be determined (up to a constant) by integrating with respect to $(\tilde{U},\tilde{B})$:

\begin{equation}
\begin{split}
	\Phi_{\tilde{H}} \left( \tilde{U}, \tilde{B} \right) = \sum_i \int_0^{\tilde{U}} \text{d}U' \langle n_{i\uparrow} n_{i\downarrow} \rangle_{\tilde{H}} ( U',0 )\\
	\quad - \sum_i \int_0^{\tilde{B}} \text{d}B' \langle S_i^z \rangle_{\tilde{H}} ( \tilde{U},B' ) + \Phi_{\tilde{H}}(0,0)
\end{split} \label{eqn:FreeEnergyIntegration}
\end{equation}
The constant $\Phi_{\tilde{H}} (0,0)$ corresponds to the grand canonical potential of a tight-binding model, which may be evaluated analytically. However, this constant is not relevant when searching for the minima of $\tilde{\Phi}$.
\section{Results}\label{sec:Results}

\subsection{Benchmarking: Four-Site Model}\label{ssec:Foursite}
To assess merit and shortcomings of the variational method suggested, here, we perform benchmark calculations for a system which can also be solved exactly. In the following, we compare solutions of the Hubbard-Heisenberg Hamiltonian, Eq.(\ref{eqn:OriginalHamiltonian}), on a 4-site cluster obtained with exact diagonalization, the Hartree-Fock approximation and with the generalized variational principle explained in Sec.(\ref{ssec:VariationalPrinciple}).

\begin{figure}[htp!]
	\centering
	\includegraphics[width=0.5\textwidth]{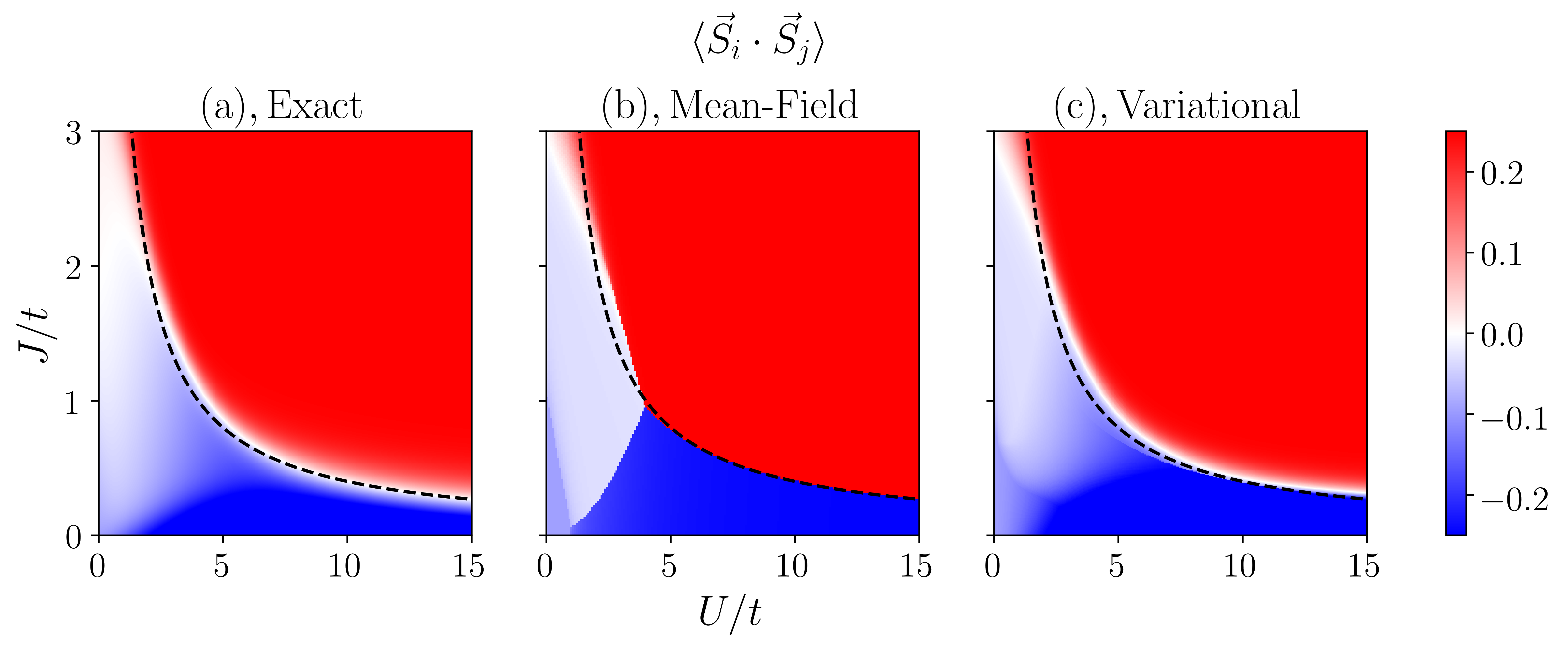}
	\caption{Spin-spin correlation between next neighbors on a 4-site model. The $4t^2/U$-line is where the transition is expected analytically in the strong-$U$ limit. (a): Exact diagonalization. (b): Mean-field treatment which allows ferromagnetic and antiferromagnetic ordering. (c): Variational approach. }\label{4SiteSS}
\end{figure}

We treat the system at half filling by setting the chemical potential to $\mu = U/2$. Fig.(\ref{4SiteSS}) then shows the total spin-spin correlation $\langle \vec{S}_i \cdot \vec{S}_j \rangle$ obtained by the three different approaches: exact diagonalization, a spin-unrestricted mean-field treatment which allows for both ferromagnetic and antiferromagnetic solutions (see Appendix (\ref{sec:AppdxMeanField})) and the variational approach which uses the effective Hamiltonian $\tilde{H}$ (Eq.(\ref{eqn:EffectiveHamiltonian})). Before analyzing the data, it is important to note that for the strong-coupling case ($U \gg t$), the behaviour of the system is known, as the Hubbard model becomes equivalent to a Heisenberg model \cite{FromHubbardToHeisenberg} with an antiferromagnetic kinetic exchange coupling $-4t^2/U$ between nearest-neighbor spins. In our case, the kinetic exchange competes with the ferromagnetic direct exchange $J$ and one obtains:
\begin{equation*}
	H \overset{U \gg t}{\approx} - \left( J - \frac{4t^2}{U} \right) \sum_{\langle i,j \rangle} \vec{S}_i \cdot \vec{S}_j
\end{equation*}
From this, it is easy to see that the nearest-neighbor spin-spin correlations should change sign at $J = 4t^2/U$ in the large-$U$ limit. A $4t^2/U$-line is plotted as a dashed, black curve inside the pictures. 

The exact solution in Fig.(\ref{4SiteSS}a) shows continuous transitions from positive to negative nearest-neighbor spin-spin correlations. As expected, antiferromagnetic correlations dominate where $J$ is small, while a large $J$ leads to ferromagnetic correlations. One should keep in mind, however, that the finite size of the system prohibits actual antiferromagnetic or ferromagnetic ordering in the exact solution. This differs from the other approximative cases where symmetry breaking is explicitly allowed. The MFT treatment in Fig.(\ref{4SiteSS}b) correctly predicts a competition between direct and kinetic exchange in the strong-$U$ limit, however fails to capture the correct order of the transition as the system undergoes a first-order phase transition from the Néel to the ferromagnetic state (and vice versa), which does not occur in the exact solution. %This is in line with mean field treatments for $T = 0$ which also display the same first order transition in the strong-$U$ limit \cite{HubbardHeisenbergMeanFieldZeroT}.

The spin-spin correlations as calculated with the variational approach (Fig.(\ref{4SiteSS}c)) are in much closer agreement with the exact solution (Fig. (\ref{4SiteSS})a) than the MFT results (Fig.(\ref{4SiteSS}b). It should be noted, that, similarly to the mean-field result, a small step (i.e., a first order transition) is still visible for intermediate $U$, which is an artifact of the method itself. This problem is, however, much less severe than in the mean-field treatment. In the $U = 0$ case, the variational approach and the MFT become equivalent. Both yield the same result in this case, as it must be.

\begin{figure}[htp!]
	\centering
	\includegraphics[width=0.5\textwidth]{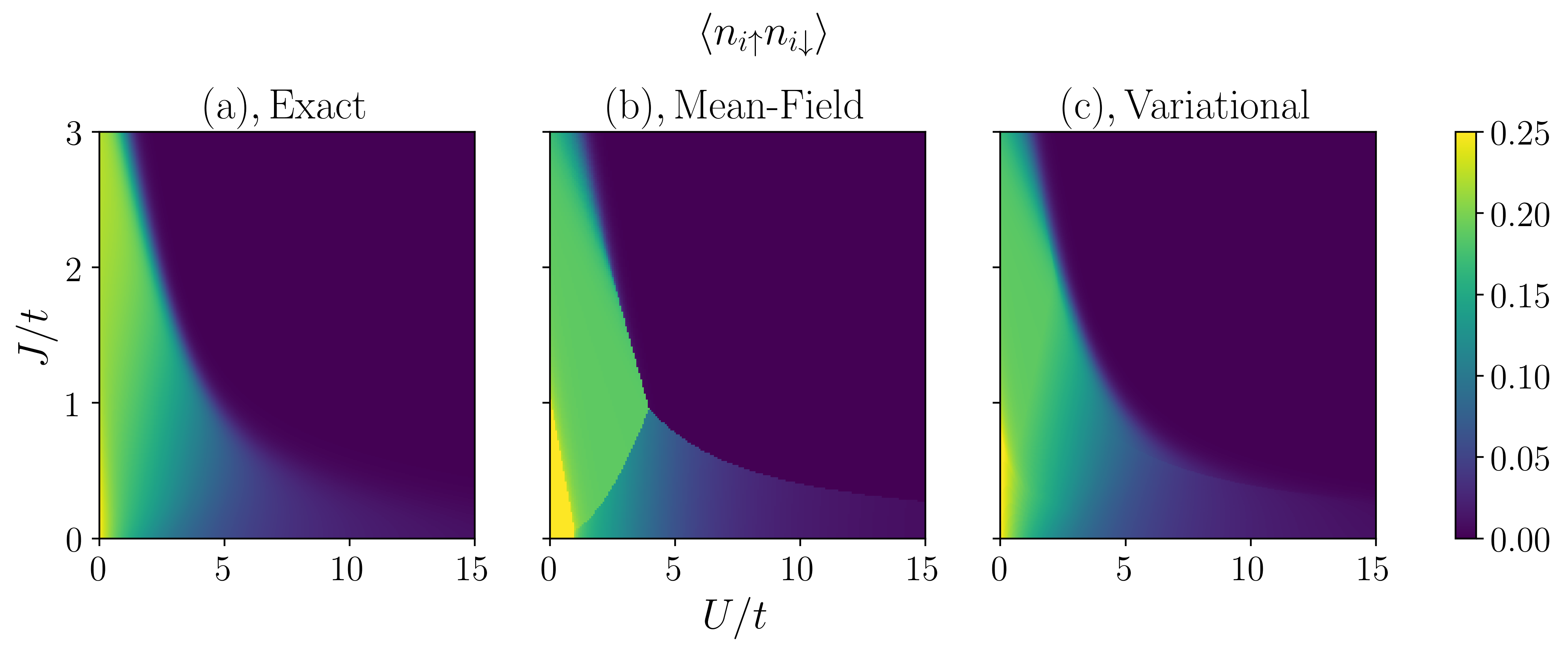}
	\caption{Double occupancy on a 4-site model. (a): Exact diagonalization. (b): Hartree-Fock Mean-Field treatment which allows ferromagnetic and antiferromagnetic ordering. (c): Variational approach.}\label{4SiteDO}
\end{figure}

The double occupancy, as obtained from the exact solution, MFT and the variational approach, is shown in Fig.(\ref{4SiteDO}. The transitions seen in the spin-spin correlation functions manifest themselves also in the $U$-$J$ dependence of the double occupancies. In the exact solution, it is visible that both the Hubbard-$U$ and the direct exchange $J$, by themselves, tend to reduce the double occupancy, and thus localize the electrons. However, kinetic (antiferromagnetic) and direct(ferromagnetic) exchange can cancel each other, which leads to non-monotonous behaviour when both interactions are present.

\begin{figure}[htp!]
	\centering
	\includegraphics[width=0.5\textwidth]{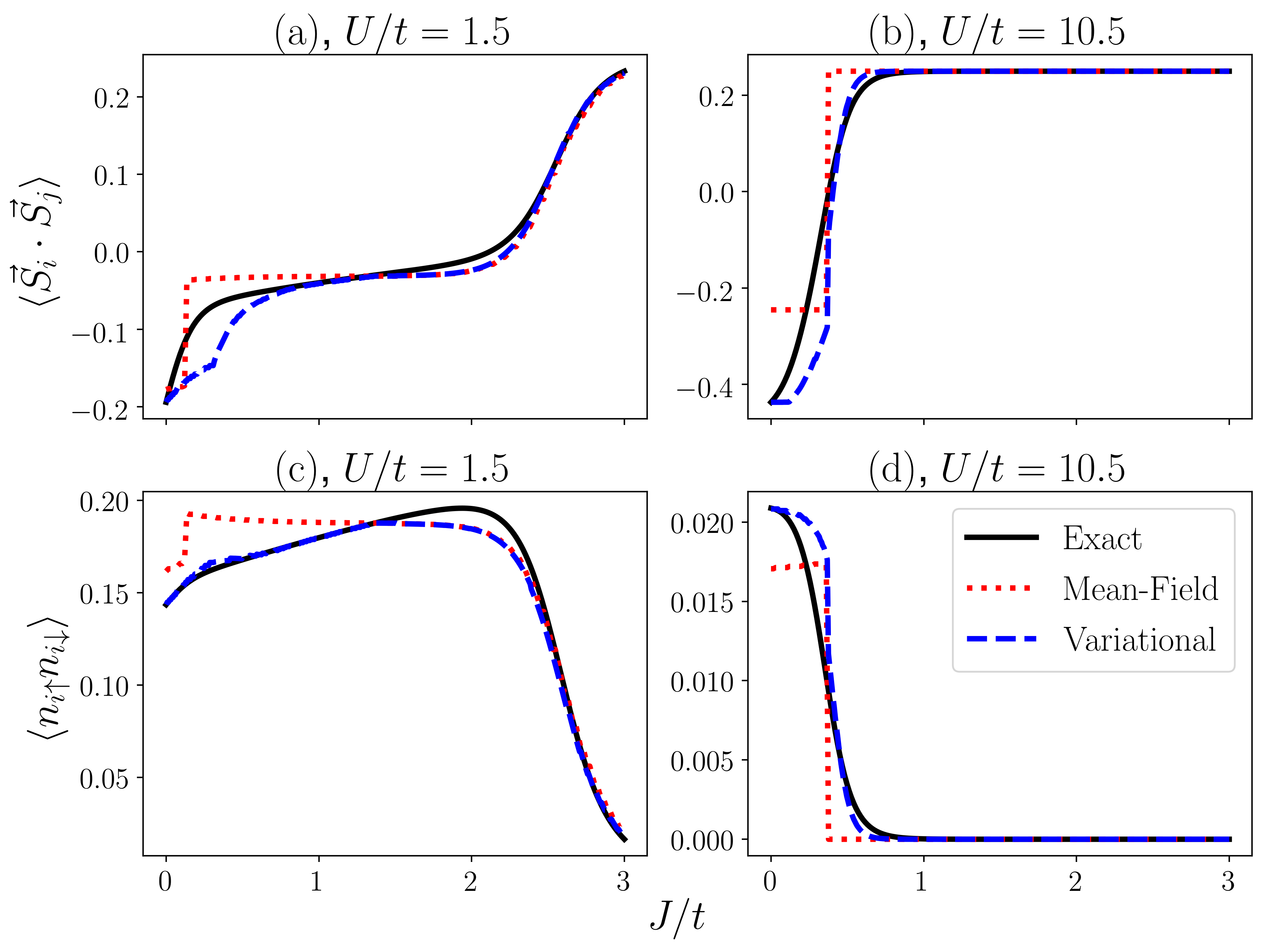}
	\caption{Correlation Functions for two different, fixed values $U$, depending on $J$. The black line shows the exact solution. The red dotted line stems from the mean-field solution, while the blue dashed lines come from our variational approach. (a),(b): Next-neighbor spin-spin correlation. (c),(d): Double occupancy.}\label{4SiteLines}
\end{figure}

This can be seen more clearly from the lineplots of the spin-spin correlation function and the double occupancy shown in Fig.(\ref{4SiteLines}). For both values of $U$, the variational approach approximates the exact solution much closer than the MFT treatment. One important thing to consider is the spin-spin correlation for a strong Hubbard-$U$ (Fig.(\ref{4SiteLines})b). For small $J$, MFT describes the antiferromagnetic correlations by explicitly going into a Néel-state, which leads to a value of $\langle \vec{S}_i \cdot \vec{S}_j \rangle = - \frac{1}{4}$ for spin$-\frac{1}{2}$ fermions (i.e. electrons). However, antiferromagnetic correlations are, as shown in the plot, much stronger. Since the auxiliary system $\tilde{H}$ is a correlated problem due to the on-site interaction $\tilde{U}$, it is capable of capturing such behaviour.

\subsection{Hubbard-Heisenberg Model on the Square Lattice}\label{ssec:SquareLattice}
In the following, we study the spin-spin correlation functions and the double occupancies of the Hubbard-Heisenberg model on the square lattice obtained in MFT and with the generalized variational approach.

\begin{figure}[h]
	\centering
	\includegraphics[width=0.5\textwidth]{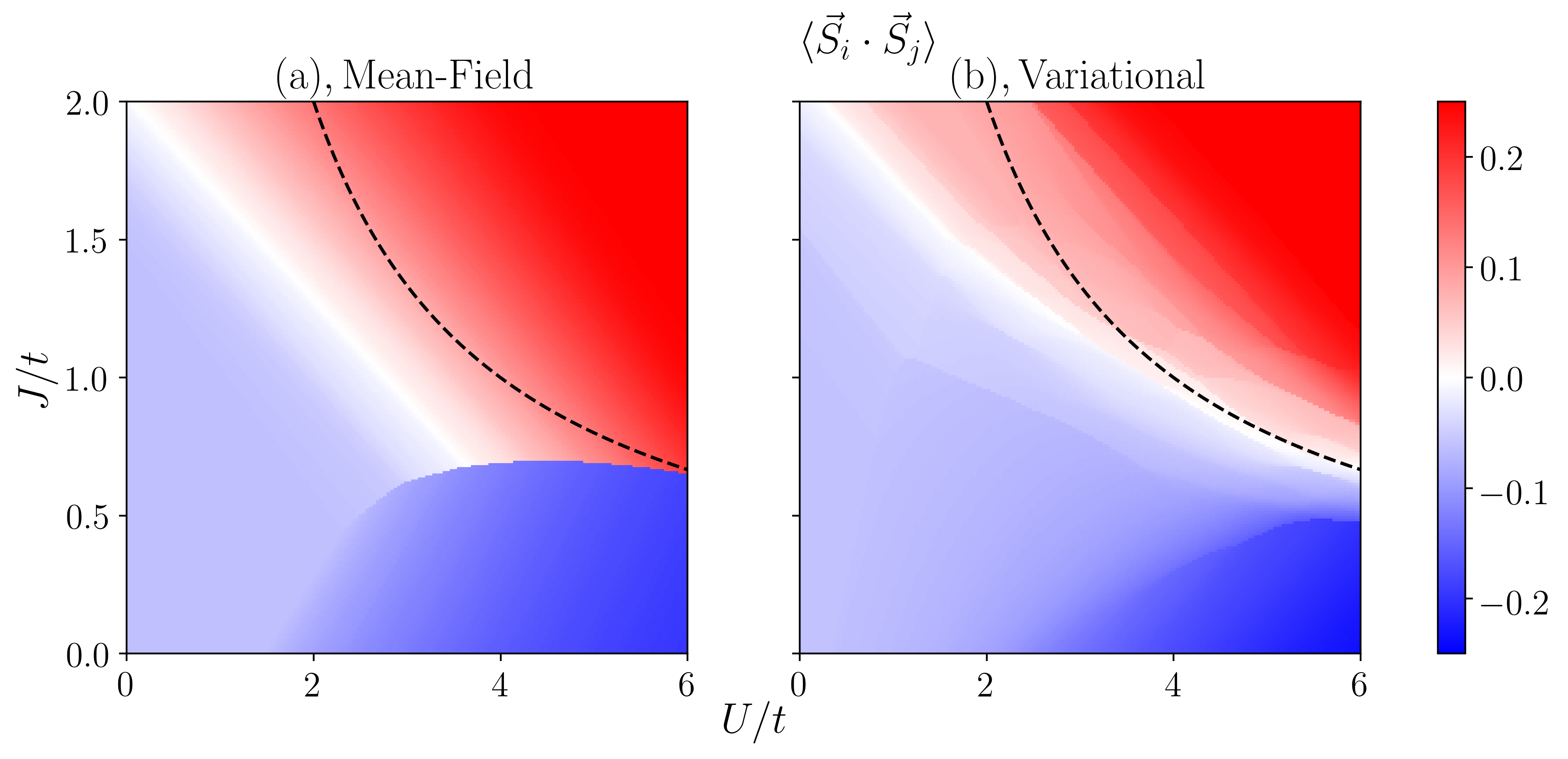}
	\caption{Spin-Spin correlation of the Hubbard-Heisenberg model between next neighbors on a half-filled square lattice. The $4t^2/U$-line is where the transition is expected analytically in the strong-$U$ limit. (a): Hartree-Fock mean-field treatment which allows ferromagnetic and antiferromagnetic ordering. (b): Variational approach.}\label{pic:squareSS}
\end{figure}

Fig.(\ref{pic:squareSS}a) shows the nearest-neighbor spin-spin correlation function, depending on $U$ and $J$ obtained from MFT. Qualitatively, it behaves similarly to the mean-field solution of the 4-site model, where the transition from dominantly antiferromagnetic to ferromagnetic spin-correlations are continuous for small $U$, while a discontinuity appears with increasing $J$ for larger $\tilde{U} \gtrsim 3$. The transition does not occur at the $4t^2/U$ line in the parameter regime of $U<6$ shown, here. At stronger on-site repulsion, $U>7$ MFT does however recover the transition at $J \approx 4t^2/U$ as expected in the strong coupling limit (see Sec.(\ref{sec:AppdxMeanField})). 

Fig.(\ref{pic:squareSS}b) shows the $U$- and $J$-dependence of the spin-spin correlation function, obtained from the variational principle. The spin-spin correlation functions obtained within mean-field theory and with the variational principle are similar regarding the global shape of predominantly ferromagnetically and antiferromagnetically correlated regions. Differences occur however at a quantitative level. First, in the variational principle the crossover point $\langle \vec{S}_i \cdot \vec{S}_j \rangle = 0$ from antiferro- to ferromagnetic correlations approaches the strong-coupling expectation of $J = 4t^2/U$ already at much smaller on-site interaction $U / t \approx 3$ than in the MFT case. Furthermore, the $J$-induced crossover from predominant antiferromagnetic to ferromagnetic spin-correlations is smoother than in MFT. While several smaller steps are still appearing in the spin-spin correlation function calculated with the variational principle, we find that these steps are within our estimated errors (see Appendix (\ref{ssec:AppdxErrorJumps})).

Furthermore, a step is still visible even at $U = 0$, where our results should coincide with the MFT. Since there is no systematic Trotter-error for $U = 0$, this clearly hints that the steps stem from finite size effects which remain even after extrapolating to $N \rightarrow \infty$. This leads us to the conclusion that, within our variational framework, the transition into the area with ferromagnetic correlations with increasing $J$ is smooth.

\begin{figure}[h]
	\centering
	\includegraphics[width=0.5\textwidth]{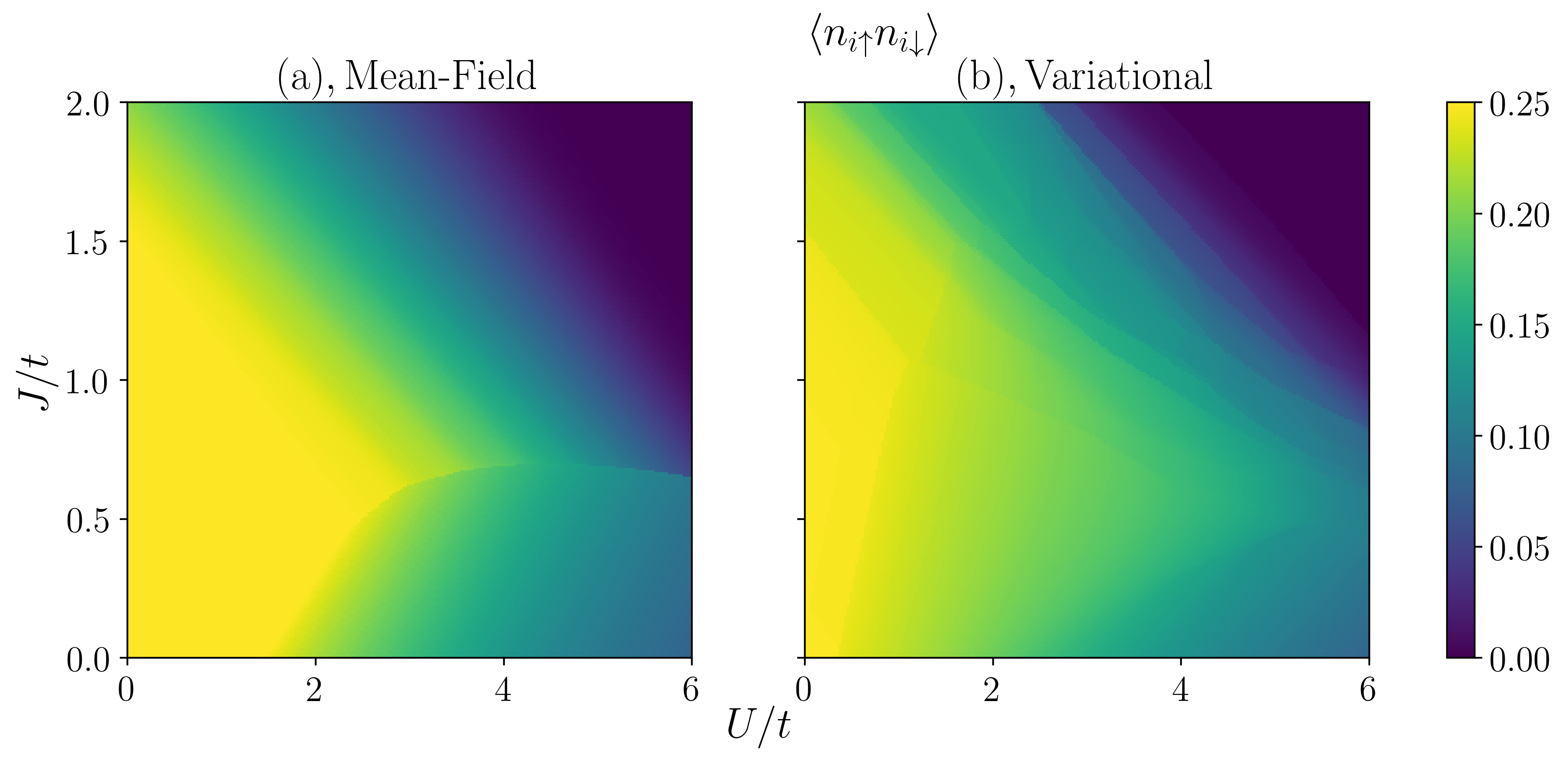}
	\caption{Double occupancy of the Hubbard-Heisenberg model on a half-filled square lattice. (a): Hartree-Fock mean-field treatment which allows ferromagnetic and antiferromagnetic ordering. (b): Variational approach.}\label{pic:squareDO}
\end{figure}

\begin{figure}[h]
	\centering
	\includegraphics[width=0.4\textwidth]{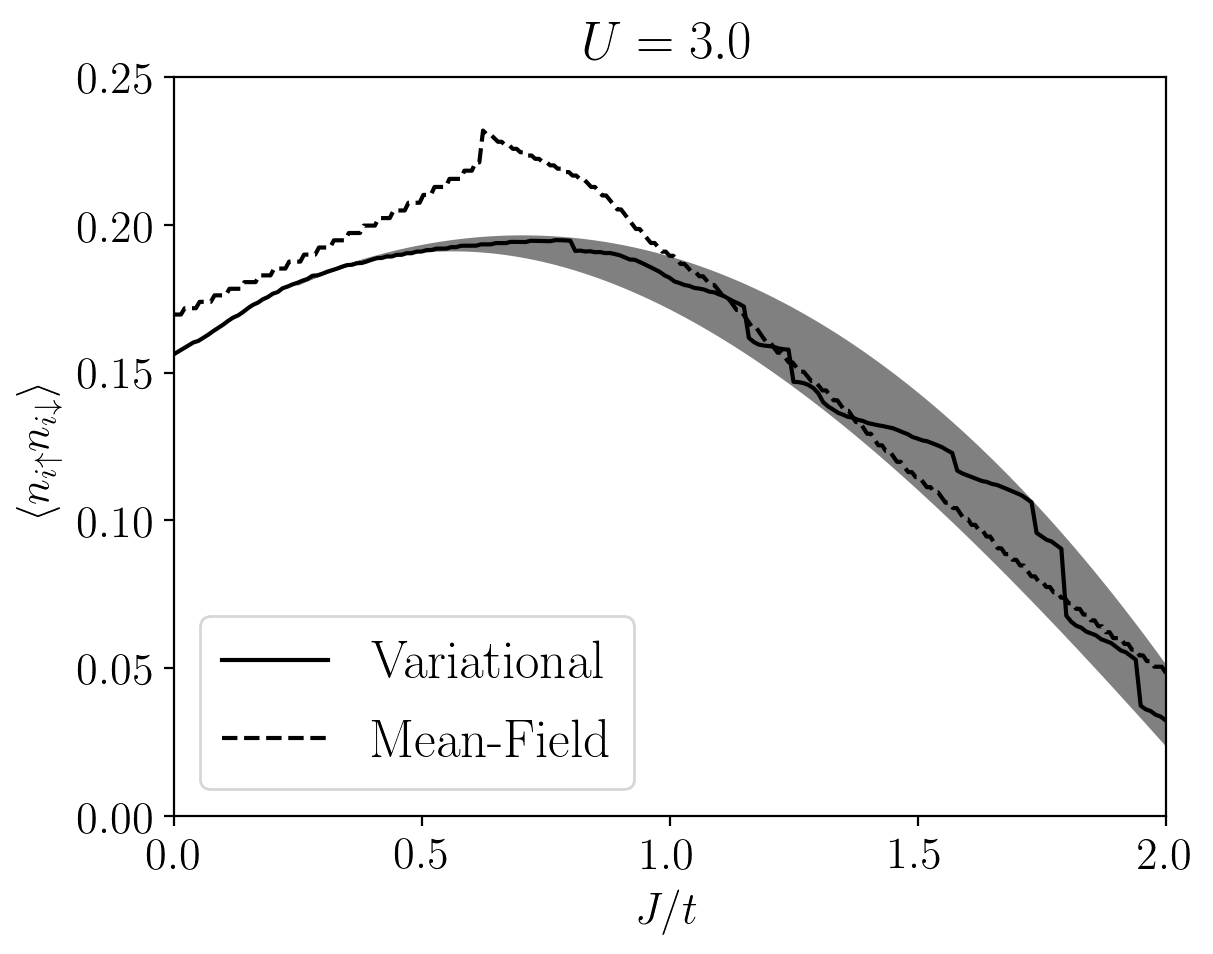}
	\caption{Double occupancy pf the Hubbard-Heisenberg model as function of nearest-neighbor Heisenberg exchange $J$ at a fixed $U/t = 3$. The shaded, grey area is meant to mark uncertainties resulting from finite size effects in the DQMC data.}\label{pic:squareDOline}
\end{figure}

Fig.(\ref{pic:squareDO}) shows the double occupancy in MFT (a) and within the variational framework (b) depending on $U$ and $J$. Qualitatively, as is the case for the spin-spin correlations, both results are roughly similar. However, the transition between states with different expectation values of the double occupancies is smooth within the variational approach, whereas MFT gives again a discrete step for $U/t \gtrsim 3$. Noteably, as illustrated in a line plot in Fig.(\ref{pic:squareDOline}), the influence of $J$ on the double occupancy is non-monotonous, which we also observed for the 4-site cluster (see Fig.(\ref{4SiteDO})). While both the $J$-term and the $U$-term in $H$ support the formation of local magnetic moments and thus reduce the double occupancy, the interplay between the two can lead to a higher value. At $U/t = 3$ in Fig.(\ref{pic:squareDOline}), where the Mott-Heisenberg picture (competition between direct exchange $J$ and kinetic exchange $-4t^2/U$) is not appropriate, this can be understood as a competition between Heisenberg-type ferromagnetism and Slater-type antiferromagnetism. A $J$ which is of the order of the hopping amplitude $t$ can lead to a non-negligible difference in the double occupancy and other correlation functions.

\section{Conclusions and Outlook}
In this work, we investigated the properties of a Hubbard-Heisenberg model, which interpolates between many mechanisms for magnetism, i.e. the Slater-, Stoner- and Heisenberg-picture. For the realistic $SU(2)$-case, we presented the first $U$-$J$ phase diagram for a half-filled square lattice which goes beyond static mean-field theory, by employing a Hubbard model with broken spin symmetry as an effective Hamiltonian through the use of the Feynman-Peierls-Bogoliubov variational principle. While both interactions present in the system lead to the formation of local magnetic moments, the interplay between the two (e.g. the competition between Slater-type antiferromagnetism and Heisenberg-type ferromagnetism) can lead to non-monotonous behaviour in properties such as the double occupancy. Compared to the discrete transitions between areas with dominant antiferromagnetic and ferromagnetic correlations, obtained within the Hartree-Fock mean-field treatment, the variational approach leads to continuous transitions.\\
\begin{center}
	\textbf{\textit{Acknowledgements}}
\end{center}
This work has been performed within the research program of the DFG Research Training Group \textit{Quantum Mechanical Materials Modeling} (QM$^3$) (Project P3). We thank Erik van Loon for many insightful discussions. The authors furthermore acknowledge the North-German Supercomputing Alliance (HLRN) for providing resources and the computing time necessary for carrying out the DQMC simulations.

\clearpage
\appendix
\section{Mean-Field solutions}\label{sec:AppdxMeanField}
The mean-field solutions for the 4-site model and the square lattice can in principle be obtained by the decoupling of the interaction terms in the original Hamiltonian (Eq.(\ref{eqn:OriginalHamiltonian})). Here, however, we use the variational principle (which leads to a completely analogous solution) by employing a non-interacting, effective Hamiltonian $\tilde{H}$ which allows both for ferro- and antiferromagnetism through two effective fields:
\begin{equation}
\begin{split}
	\tilde{H} = - t \sum_{\langle i,j\rangle , \sigma} \left( a^{\dagger}_{i\sigma} b^{\vphantom{\dagger}}_{j\sigma} + h.c. \right) - \tilde{B}_1 \sum_i \left( S_{ia}^z + S_{ib}^z \right)\\ - \tilde{B}_2 \sum_i \left( S_{ia}^z - S_{ib}^z \right)
\end{split}
\end{equation}
Noteably, the staggered magnetic field with the magnitude $\tilde{B}_2$ breaks translational symmetry due to the induced Néel-order. Hence, the original square lattice is divided into two sublattices, leading to two distinct fermionic operators for the respective sublattices. The parameters ($\tilde{B}_1$,$\tilde{B}_2$) are chosen variationally for each set of original ($U$,$J$).

\begin{figure}[h]
	\centering
	\includegraphics[width=0.5\textwidth]{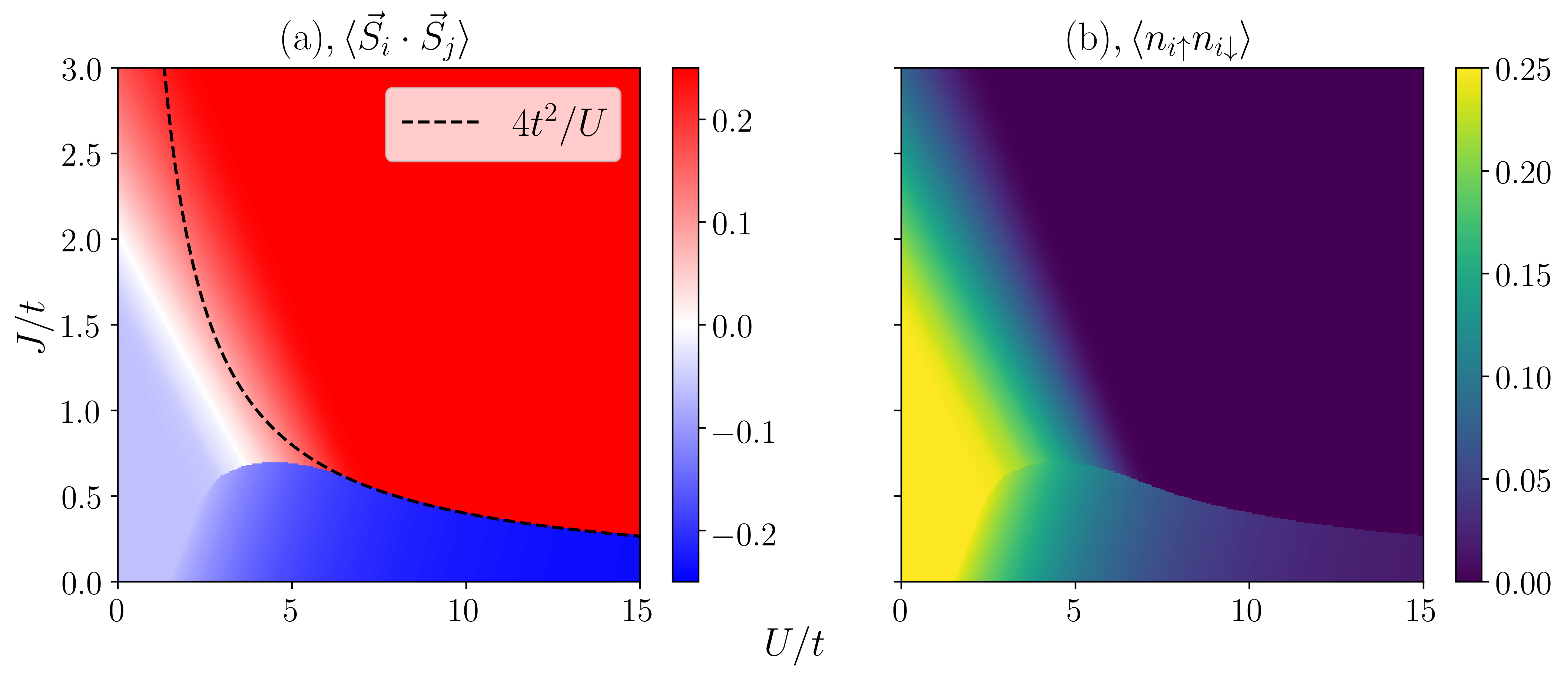}
	\caption{Mean-field solutions for a half filled square lattice. (a): Next-neighbor spin-spin correlation with the $4t^2/U$ line. (b): Double occupancy.}\label{pic:MFresultsAppdx}
\end{figure}

We can solve the effective, non-interacting Hamiltonian analytically through simple fourier transform of the fermionic operators and find (with the lattice constant set to $a = 1$) the following four bands: 
\begin{equation*}
	\varepsilon_{1,2}^{\sigma} (\vec{k}) = - \frac{\tilde{B}_1 \sigma}{2} \pm \sqrt{ \left(\frac{ \tilde{B}_2}{2} \right) ^2 + \left( 4 t \cos (\frac{k_x}{\sqrt{2}}) * \cos (\frac{k_y}{\sqrt{2}}) \right) }
\end{equation*}
From this, all relevant expectation values can be computed exactly, either directly through the derivatives of the grand potential or through Wick factorization. If we then write out the variational equation (Eq.(\ref{eqn:VariationGeneral})) explicitly, we obtain the following expression which needs to be minimized with respect to ($\tilde{B}_1$,$\tilde{B}_2$):
\begin{equation}
\begin{split}
	\tilde{\Phi} = U \sum_i \langle n^a_{i\uparrow}n^a_{i\downarrow} + n^b_{i\uparrow}n^b_{i\downarrow} \rangle_{\tilde{H}} - J \sum_{\langle i,j \rangle} \langle \vec{S}_i^a \cdot \vec{S}_j^b \rangle_{\tilde{H}} \\ + \tilde{B}_1 \sum_i \langle S_{ia}^z + S_{ib}^z \rangle_{\tilde{H}} + \tilde{B}_2 \sum_i \langle S_{ia}^z - S_{ib}^z \rangle_{\tilde{H}} + \Phi_{\tilde{H}} 
\end{split}
\end{equation}
As mentioned above, employing a non-interacting effective $\tilde{H}$ within the variational framework is completely analogous to performing a decoupling of $H$ which allows for ferro- and antiferromagnetic solutions. Fig.(\ref{pic:MFresultsAppdx}) shows, again, the spin-spin correlation and double occupancy, for a greater parameter range than in Sec.(\ref{sec:Results}). The transition between Néel- and ferromagnetic order, which is analytically expected at the $J = 4t^2/U$-line in the strong-$U$ regime, can be seen clearly.

\section{Trotter Extrapolation}\label{sec:AppdxTrotter}
In this section, we provide an example in order to justify the Trotter extrapolation which was performed with only two data points ($\Delta \tau = 0.1$ and $0.2$). The error is known to scale with $\mathcal{O} ( \Delta \tau ^2 )$. Consider again the exactly solvable four-site model which was treated in Sec.(\ref{ssec:Foursite}), for the specific case of $J = 0$, i.e. a four-site Hubbard model at half filling. In order to compare to the exact results obtained from exact diagonalization, we simulated the model within DQMC with the aforementioned Trotter discretizations. The small size of the system allows for much longer samplings with 10000 warmup sweeps and 150000 measurement sweeps. The simulations were performed with $U = 0-6$, at 100 equidistant data points. Furthermore, the obtained data is smoothed with a Savitzky-Golay filter using third order polynomials and a window length of $w = 7$ data points.

\begin{figure}[h]
	\centering
	\includegraphics[width=0.5\textwidth]{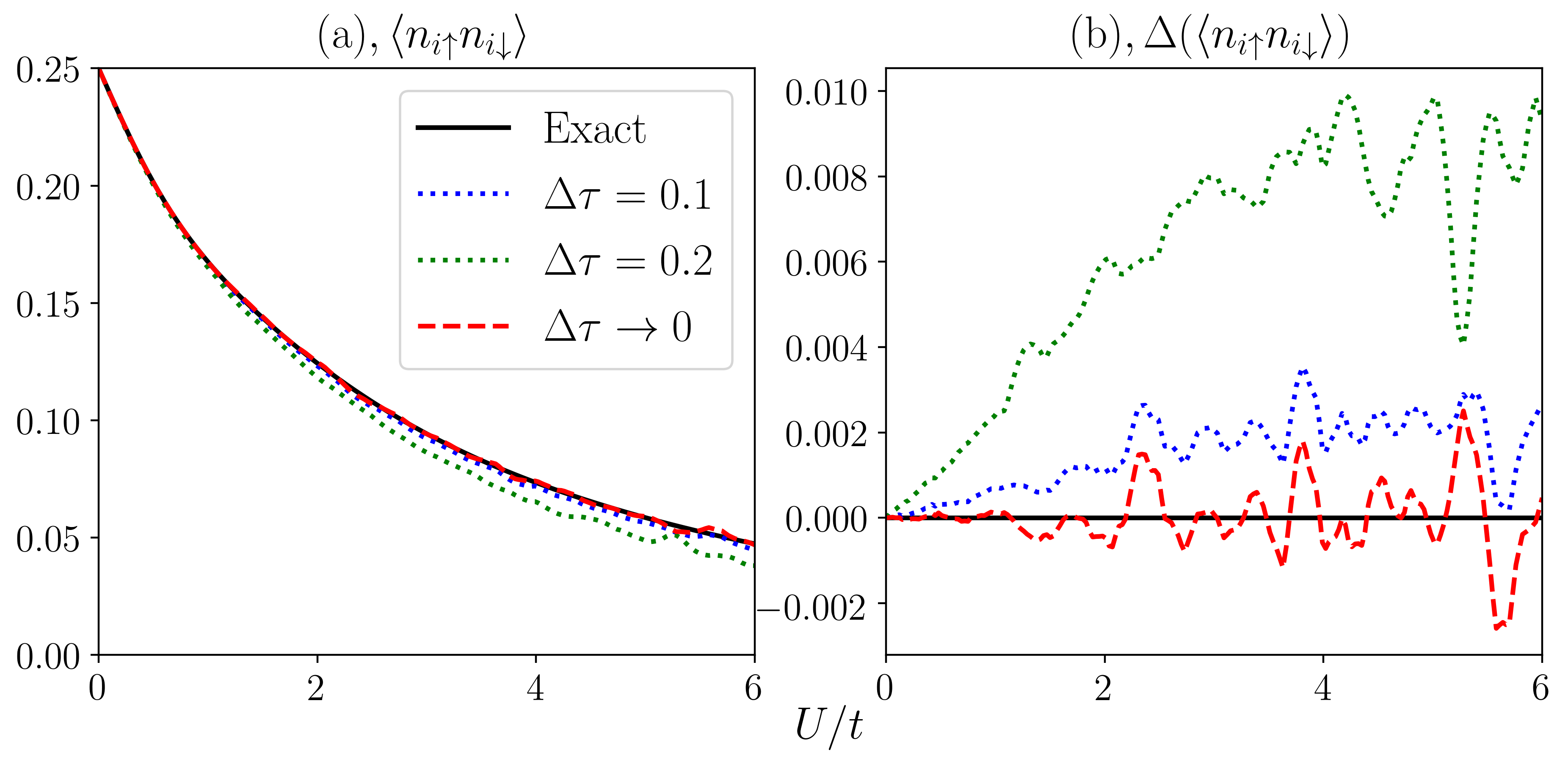}
	\caption{(a): Double occupancy of a four-site Hubbard model at half filling, obtained from ED, DQMC for two different Trotter steps and after extrapolation $\Delta \tau \rightarrow 0$. (b): Total error in the double occupancy.}\label{pic:TrotterAppdx}
\end{figure}
Fig.(\ref{pic:TrotterAppdx}a) shows the double occupancy of the model depending on $U$, obtained from ED, DQMC for both Trotter steps and the result from the extrapolation $\Delta \tau \rightarrow 0$. It is visible that the extrapolation brings the DQMC results very close to the exact solution. In order to better quantify this, Fig.(\ref{pic:TrotterAppdx}b) shows the error in the double occupancy for both discretizations. After extrapolating, the systematic error is diminished, and only the statistical noise remains, which is visible as the extrapolated curve oscillates slightly around $0$.
\section{Error Estimation}
The calculations in this work are prone to a variety of different error sources, which need to be adressed separately. First, we discuss the variational principle itself, where the exact benchmarking data provides some insight. Secondly, we turn our attention to the results obtained from DQMC, i.e. the statistical error, the finite size extrapolation and the integration errors which occur when computing the free energy.

\subsection{Systematic and Statistical Errors of DQMC} \label{ssec:AppdxErrorsDQMC}
The most obvious error source when using Monte Carlo methods is the statistical error due to a finite number of samplings. As mentioned before, we carried out simulations on an equidistant 48x48 grid for $\tilde{U}/t = 0-10$ and $\tilde{B}/t = 0-4$, with 10000 warmup and 30000 measurement sweeps. The error can be drastically reduced by making use of an appropriate filter. Similarly to another work \cite{MalteThermodynamicsMITExtendedHubbard} done in our group which also relies on DQMC, we make use of a two-dimensional Savitzky-Golay filter \cite{SavitzkyGolay} which, in a box width of $w_{\tilde{U}}$ and $w_{\tilde{B}}$, fits a two-dimensional polynomial of the form
\begin{equation*}
p \left( \tilde{U},\tilde{B} \right) = \sum_{mn}^{NM} c_{nm} \tilde{U}^n \tilde{B}^m
\end{equation*}
to the data. The polynomials are of third order, and box widths are both set to $w = 1.0$. Additionally, data which is close to the original starting point $(\tilde{U}_0,\tilde{B}_0)$, is given additional weight through a tricubic weighting function $(1 - d^3)^3$ where the distance $d$ is defined as $d = \text{max} \left\lbrace |\tilde{U} - \tilde{U}_0|/w_{\tilde{U}}, |\tilde{B} - \tilde{B}_0|/w_{\tilde{B}} \right\rbrace$.

Another important error source to address is the integration procedure (see Eq.(\ref{eqn:FreeEnergyIntegration})) when computing the grand potential $\Phi_{\tilde{H}}$ of the effective system. While the remaining statistical errors tend to cancel itself out when integrating, the remaining Trotter-error might be magnified again.

\begin{figure}[h]
	\centering
	\includegraphics[width=0.5\textwidth]{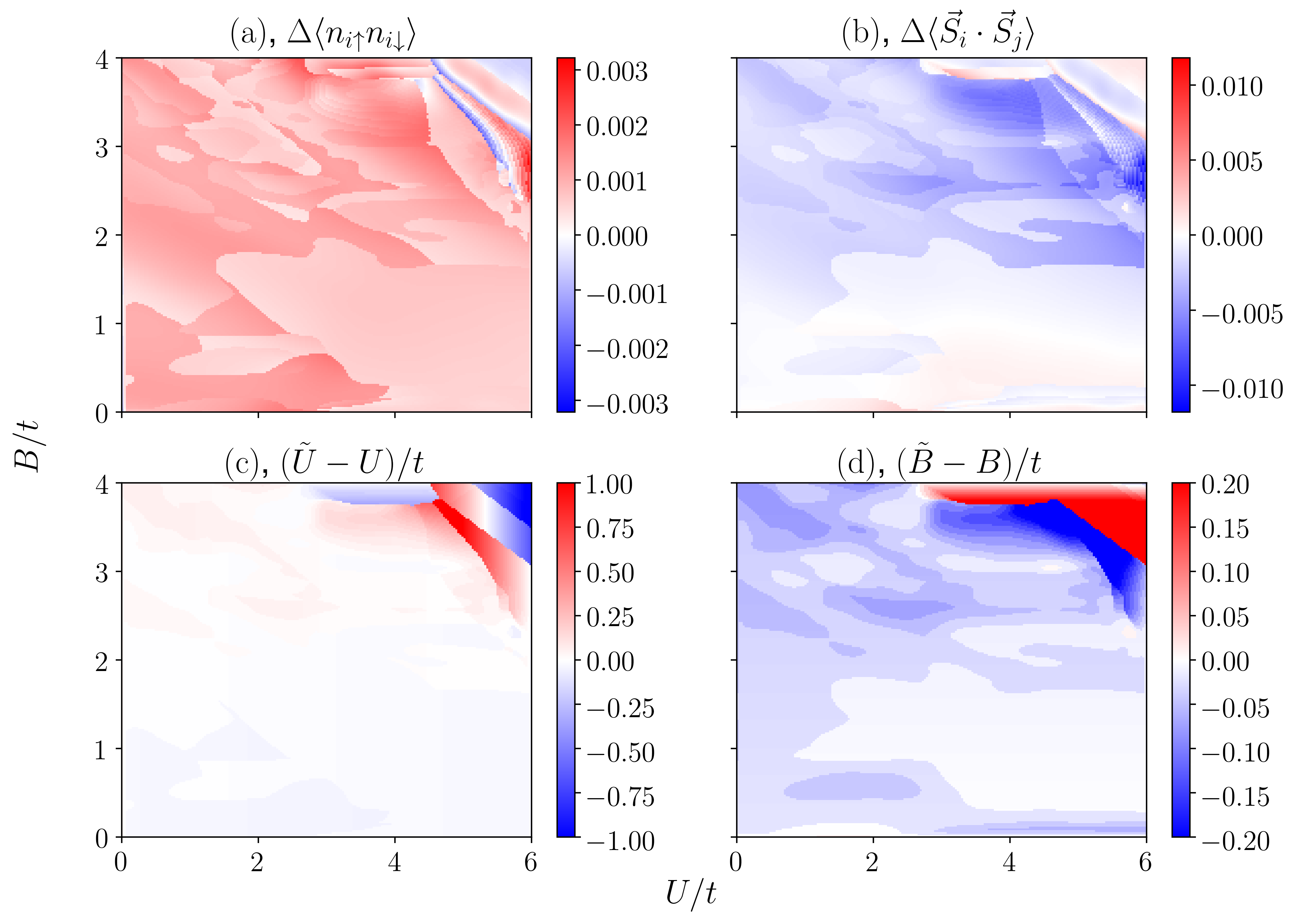}
	\caption{Absolute errors which stem from the integration procedure and the remaining Trotter error. (a): Double occupancy.,(b): Spin-spin correlation,(c): On-site interaction,(d): Magnetic field.}\label{pic:SystematicErrorsAppdx}
\end{figure}

In order to estimate the errors of the whole mapping procedure, we set $H = \tilde{H}$, mapping the effective Hamiltonian to itself, $(U,B) \rightarrow (\tilde{U},\tilde{B})$. Obviously, in the absence of errors, the parameters $(U,B)$ should not change. Fig.(\ref{pic:SystematicErrorsAppdx}) shows both the absolute errors of $(\tilde{U},\tilde{B})$ and the correlation functions (i.e. the double occupancy and the spin-spin correlation) which are necessary for the mapping.
\subsection{Finite Size Error of DQMC}
As mentioned above, we carried out simulations of the Hamiltonian in Eq.(\ref{eqn:EffectiveHamiltonian}) on a square lattice with periodic boundary conditions, with system sizes of $L = 4,6,8,10,12$ lattice sites in one direction. The data is then extrapolated to $N \rightarrow \infty$ by plotting it against $1/N^2$ (with $N=L^2$ being the total number of lattice sites) and fitting a line.

In order to assess the errors which remain after the extrapolation, it should be noted that finite size effects will be the strongest when the system is free of interactions, i.e. $\tilde{U} = 0$, since the Hubbard-$U$ generally tends to localize the electrons. Furthermore, in this case, the error from the Trotter-decomposition is non-existent since there is no interaction which needs to be decoupled. The non-interacting case, where the finite size error is the most severe, can be solved analytically, which allows us to compare our data to the exact solution.
\begin{figure}[h]
	\centering
	\includegraphics[width=0.5\textwidth]{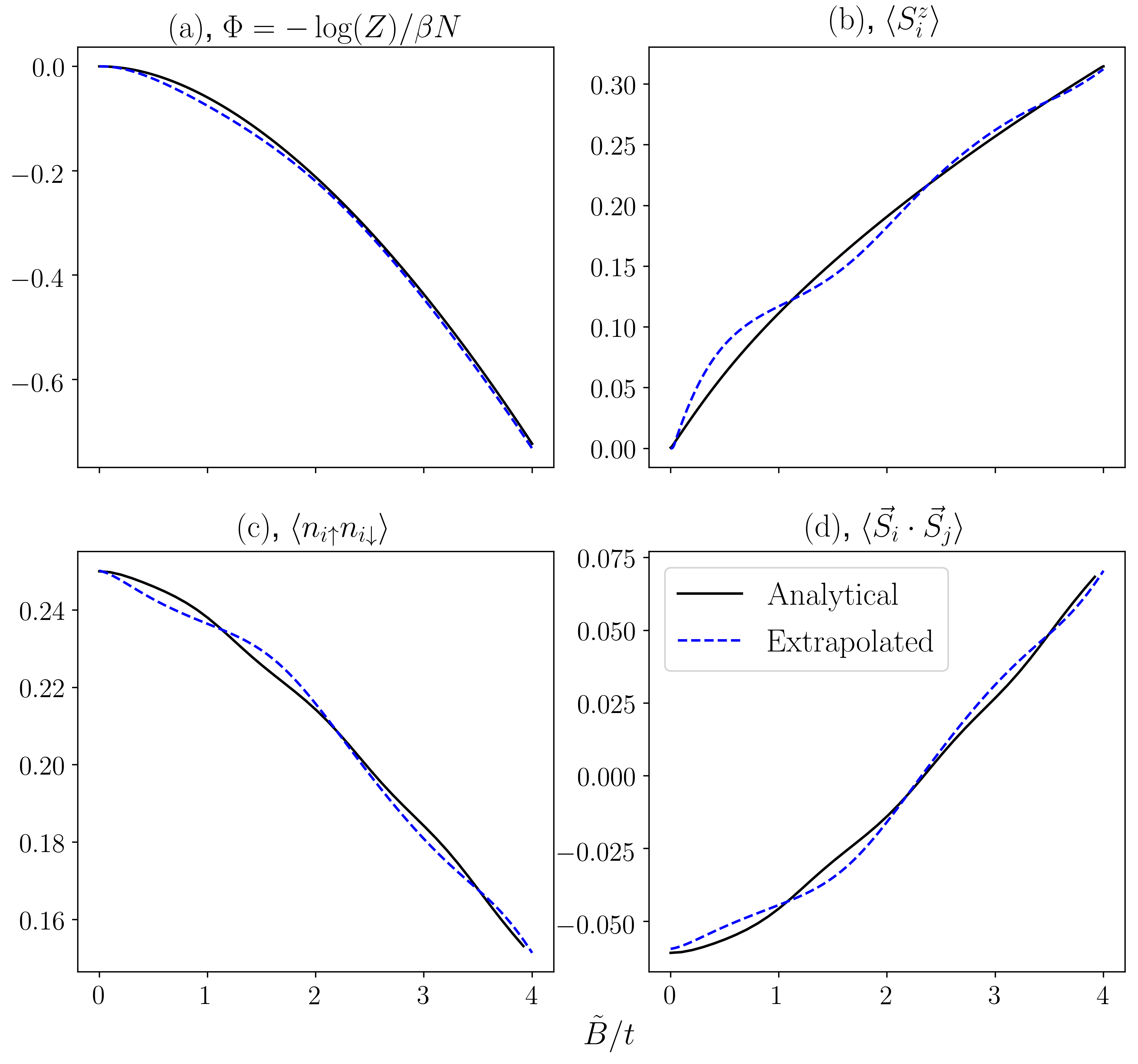}
	\caption{(a): Analytical (black line) and extrapolated (blue, dashed line) Grand Potential $\Phi$ of $\tilde{H}$ with $\tilde{U} = 0$.  (b): Magnetization, computed as the derivative of $\Phi$ with respect to $\tilde{B}$. (c): Double occupancy. (d) Next-neighbor spin-spin correlation.}\label{pic:FiniteSizeErrorsAppdx}
\end{figure}

Fig.(\ref{pic:FiniteSizeErrorsAppdx}a) shows the exact grand potential (per lattice site) of $\tilde{H}$ on a half-filled square lattice and the grand potential obtained by integrating over the smoothed and extrapolated DQMC data. Qualitatively, they are in very good agreement. However, when computing observables like the magnetization (Fig.(\ref{pic:FiniteSizeErrorsAppdx}b)) by calculating the derivatives, the small, remaining oscillations become clearly visible.

\subsection{Error estimation at the steps}\label{ssec:AppdxErrorJumps}
While there are clear hints that the steps are an artifact which stems from remaining finite size effects, we additionally demonstrate that the steps are within our estimated error bars. As a representative example, we pick the step which is visible in Fig.(\ref{pic:squareSS}b) at $U/t \approx 2.0$ and $J/t \approx 0.95$. The step itself stems from the existence of two minima in the functional $\tilde{\Phi}$, which are located at $(\tilde{U},\tilde{B})_1 \approx (1.11,0.34)$ and $(\tilde{U},\tilde{B})_2 \approx(1.02,0.88)$.

\begin{figure}[h]
	\centering
	\includegraphics[width=0.5\textwidth]{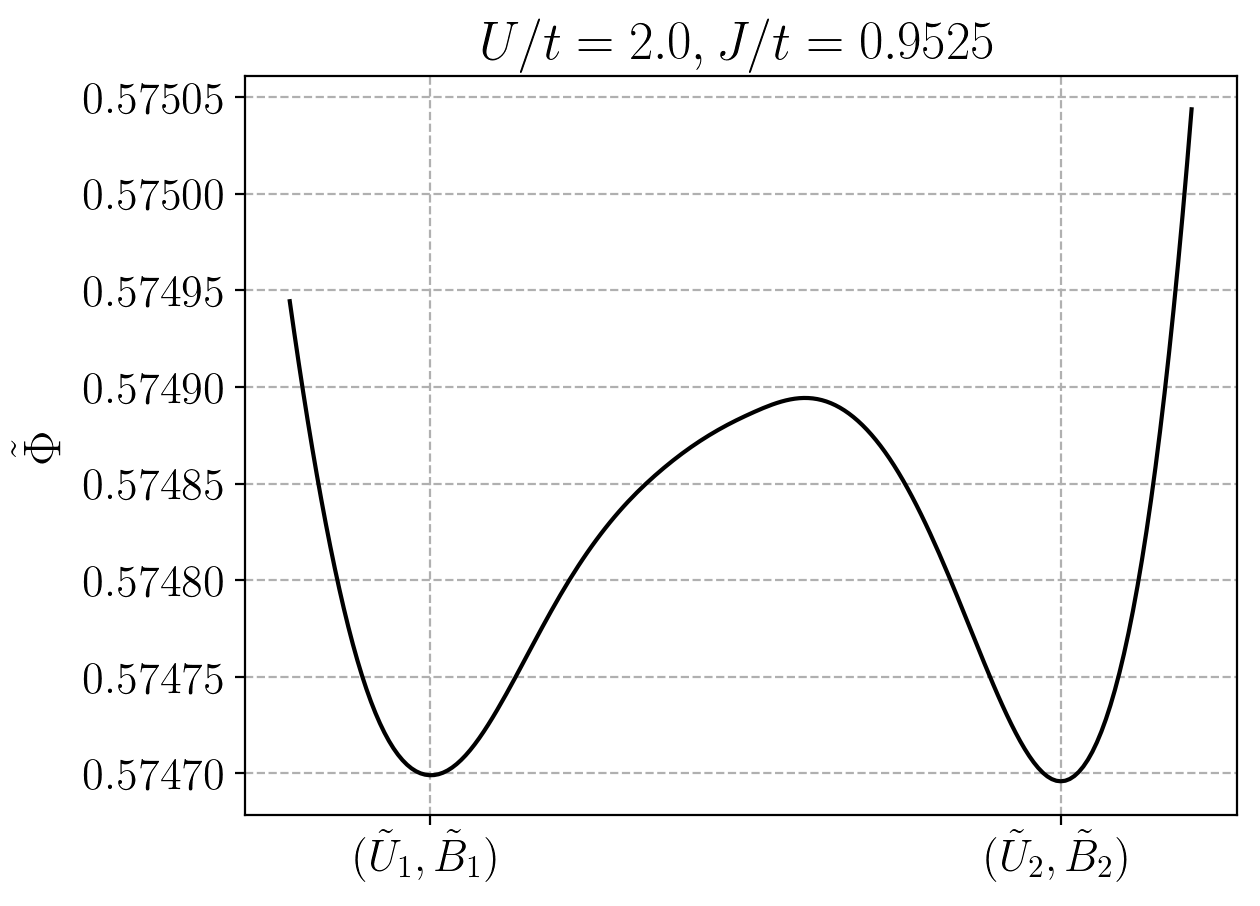}
	\caption{Functional $\tilde{\Phi}$, along the line connecting the two minima at $(\tilde{U},\tilde{B})_1 \approx (1.11,0.34)$ and $(\tilde{U},\tilde{B})_2 \approx(1.02,0.88)$, for fixed $(U,J) = (2.0,0.9525)$.}\label{pic:MinimaAppdx}
\end{figure}

Fig.(\ref{pic:MinimaAppdx}) shows a lineplot of the functional $\tilde{\Phi}$ in the $(\tilde{U},\tilde{B})$-plane along the line connecting two minima. The potential barrier between the minima has a height of $h_{\tilde{\Phi}} < 0.0002$. However, we estimate that the error for $\tilde{\Phi}$ is on a scale $\Delta \tilde{\Phi} \sim 0.05 $, which is clearly larger than the barrier height. Thus, the transition is smooth within our error bars.

\bibliography{paper.bib}
\end{document}